# MEG abnormalities and mechanisms of surgical failure in neocortical epilepsy

Thomas W. Owen[1] | Gabrielle M. Schroeder[1] | Vytene Janiukstyte[1] | Gerard R. Hall[1] | Andrew McEvoy[2] | Anna Miserocchi[2] | Jane de Tisi[2] | John S. Duncan[2] | Fergus Rugg-Gunn[2] | Yujiang Wang[1,2,3] | Peter N. Taylor[1,2,3]

[1]Computational Neurology, Neuroscience & Psychiatry Lab, ICOS Group, School of Computing, Newcastle University, Newcastle upon Tyne, UK

[2]UCL Queen Square Institute of Neurology, London, UK

[3]Faculty of Medical Sciences, Newcastle University, Newcastle upon Tyne, UK

**Correspondence**
Thomas W. Owen and Peter N. Taylor, CNNP Lab, Interdisciplinary Computing and Complex BioSystems Group, School of Computing, Newcastle University, Newcastle upon Tyne, UK.
Email: t.w.owen1@newcastle.ac.uk; peter.taylor@newcastle.ac.uk

**Funding information**
Engineering and Physical Sciences Research Council, Grant/Award Number: EP/L015358/1; Medical Research Council, Grant/Award Number: MR/K005464/1; NIHR UCLH/UCL Biomedical Research Centre; UK Research and Innovation, Grant/Award Number: MR/T04294X/1 and MR/V026569/1; Wellcome Trust, Grant/Award Number: 218380

**Abstract**

**Objective:** Epilepsy surgery fails to achieve seizure freedom in 30%–40% of cases. It is not fully understood why some surgeries are unsuccessful. By comparing interictal magnetoencephalography (MEG) band power from patient data to normative maps, which describe healthy spatial and population variability, we identify patient-specific abnormalities relating to surgical failure. We propose three mechanisms contributing to poor surgical outcome: (1) not resecting the epileptogenic abnormalities (mislocalization), (2) failing to remove *all* epileptogenic abnormalities (partial resection), and (3) insufficiently impacting the overall cortical abnormality. Herein we develop markers of these mechanisms, validating them against patient outcomes.

**Methods:** Resting-state MEG recordings were acquired for 70 healthy controls and 32 patients with refractory neocortical epilepsy. Relative band-power spatial maps were computed using source-localized recordings. Patient and region-specific band-power abnormalities were estimated as the maximum absolute $z$-score across five frequency bands using healthy data as a baseline. Resected regions were identified using postoperative magnetic resonance imaging (MRI). We hypothesized that our mechanistically interpretable markers would discriminate patients with and without postoperative seizure freedom.

**Results:** Our markers discriminated surgical outcome groups (abnormalities not targeted: area under the curve [AUC] = 0.80, $p = .003$; partial resection of epileptogenic zone: AUC = 0.68, $p = .053$; and insufficient cortical abnormality impact: AUC = 0.64, $p = .096$). Furthermore, 95% of those patients who were not seizure-free had markers of surgical failure for at least one of the three proposed mechanisms. In contrast, of those patients without markers for any mechanism, 80% were ultimately seizure-free.

**Significance:** The mapping of abnormalities across the brain is important for a wide range of neurological conditions. Here we have demonstrated that interictal MEG band-power mapping has merit for the localization of pathology and improving our mechanistic understanding of epilepsy. Our markers for mechanisms of surgical failure could be used in the future to construct predictive models of surgical outcome, aiding clinical teams during patient pre-surgical evaluations.

## 1 | INTRODUCTION

Epilepsy surgery fails to completely suppress seizures in 30%–40% of cases.[1] However, for patients with focal neocortical epilepsy, the chance of postoperative seizure freedom is even lower.[1,2] It is not fully understood why some patients with neocortical epilepsy have unfavorable surgical outcomes, although some mechanisms have been proposed. Suggested surgical failure mechanisms include incorrect localization, incomplete resection of the epileptogenic zone, or the presence of a secondary distant epileptogenic zone.[3] Other studies suggest that the development of new epileptogenic zones postoperatively,[4] and the presence of more widespread distributed epileptogenic networks,[5] may also relate to surgical failure. Multiple mechanisms could lead to poor surgical outcome in terms of seizure freedom (considered as surgical failure), with different mechanisms likely in different patients.[6]

Although the mechanisms of surgical failure may differ between patients, a central concept is that of the epileptogenic zone—a presumably abnormal area, which is indispensable for seizure generation,[7] and thus should be targeted by surgery. To localize the epileptogenic zone and predict surgical outcomes, previous studies used preoperative resting-state magnetoencephalography (MEG) data.[8–12] These studies typically focused on patient data only, without incorporating normative data from healthy controls into the analysis. Without accounting for the normative variations in health, it is difficult to identify abnormalities. Relating abnormalities relative to health with surgical outcome has been fruitful using structural magnetic resonance imaging (MRI) data.[13–16] However, this normative approach is rarely used for functional/neurophysiological data, despite efforts to promote these methods.[17–21]

In this study, we propose three mechanisms of surgical failure, in terms of postoperative seizure relapse. These mechanisms are (1) failing to resect abnormalities, (2) partial resection of the epileptogenic zone, and (3) local resections insufficiently altering the global cortical abnormality. We develop quantitative markers for each surgical failure mechanism using interictal MEG band-power abnormality maps and investigate their potential to predict surgical outcomes.

**Key Points**

- Comparing patient magnetoencephalography (MEG) band power to healthy controls reveals patient abnormalities.
- Quantifying these abnormalities, we suggest markers of three mechanisms for surgical failure.
- Ninety-five percent of patients with postoperative seizures had at least one marker of surgical failure.
- Eighty percent of seizure-free patients had no markers of surgical failure.
- Patients may have postoperative seizures for different reasons; our study formalizes this using interictal MEG recordings.

## 2 | METHODS

### 2.1 | Patients and controls

Data were acquired pre-operatively for 32 individuals with refractory neocortical epilepsy (12 International League Against Epilepsy (ILAE) 1 and 20 ILAE 2+) and 70 healthy controls. Four patients with temporal lobe epilepsy in whom resections were confined to neocortex were included in the cohort. All patients with pre- and postoperative MRI, and pre-operative MEG acquired in London, were included. All patients later underwent neocortical surgical resection. Patient age at resection ranged between 18 and 60 years, with healthy controls selected to age match the patients (age range 19–55 years). In addition to the 32 patients, 19 patients with temporal lobe epilepsy underwent surgery for resection of the hippocampus and are included in the Supplementary Materials for completeness. Hippocampal resection patients are not included in the main text, as our MEG processing did not include deep brain structures, including the hippocampus. Surgery outcome was assessed 12 months postoperatively using the ILAE seizure-freedom scale.[22] Subject data are summarized in Table S2.5.

## 2.2 | MRI acquisition and pre-processing

T1-weighted MRI was performed using a 3T GE Signa HDx scanner. Acquisition details for patients have been reported previously[23] (acquired in London), and for healthy controls[24] (acquired in Cardiff). We include a copy of the acquisition details in the Supplementary Materials, S1.1. Subject MRI scans were pre-processed using the FreeSurfer pipeline "recon-all."[25] MRI scans were parcellated into cortical regions of interest (ROIs) based on the Lausanne parcellation for four different resolutions (68, 114, 219, and 448 neocortical ROIs).[26] To identify patient-specific resection cavities, pre- and postoperative MRI scans were linearly co-registered using the FSL tool "FLIRT."[27–29] Using FSLview, pre- and postoperative MRI were overlaid, with resection volumes manually drawn. Following this, pre- and postoperative ROI volumes were calculated using custom MATLAB code.[23] Regions were categorized as resected if the pre- and postsurgical volume change exceeded 10%. Regions with volume changes between 1% and 10% were categorized as "Unknown" and were subsequently removed from any analysis.

## 2.3 | MEG acquisition and pre-processing

Eyes-closed awake resting-state MEG recordings were acquired for patient and healthy cohorts using a 275 channel CTF whole head MEG system in a magnetically shielded room. Patient data were collected during pre-surgical evaluation at UCL in London. Healthy control normative data were collected as part of the MEG UK partnership at CUBRIC, Cardiff. Raw MEG recordings were pre-processed using Brainstorm.[30] First, MEG sensor locations and structural MRI scans were co-registered using the fiducial points with visual inspection and refinement if required. Second, MEG recordings were downsampled to 600 Hz, bandpass filtered between 1 and 100 Hz, and then filtered at 50 Hz using a second-order IIR notch filter. Cardiac and ocular artifacts were identified and removed manually using independent component analysis (ICA) after channel recordings were dimensionally reduced to 40 components using principal component analysis (PCA).

Following artifact removal, MEG data were source reconstructed using the minimum norm estimate approach, sLORETA,[31] coupled with an overlapping spheres head model. This resulted in 15 000 sources constrained perpendicular to the cortex. Sources were downsampled into neocortical ROIs based on the Lausanne parcellation.[26] Sources were downsampled independently into neocortical ROIs using each of the four Lausanne parcellation schemes.[26] We report findings for a mid-resolution parcellation (114 regions) in the main text and include results for others in the Supplementary Material for completeness. Source recordings within each ROI were sign flipped and averaged, resulting in one time series per region. Finally, time series were reduced to 70 s epochs, clear of any residual artifacts (Figure 1A). Note that deep brain subcortical volumetric structures of the Lausanne atlas were not included. Specifically, we excluded the hippocampus and amygdala, among others, from our analysis entirely. Within the main text we, therefore, present results only for patients who had resections to neocortical tissue where complete MEG coverage was present. We include patients with hippocampal resections in the Supplementary Material for completeness. Recordings were not inspected for interictal spikes here because occasional epileptic spikes have little to no effect on the regional power spectral densities and maximum abnormality estimates.[20]

## 2.4 | Normative mapping of MEG band power

Neocortical regional power spectral densities were computed using Welch's estimate, with a 2 s sliding window and 50% overlap. Absolute band power was estimated for five frequency bands: delta (1–4 Hz), theta (4–8 Hz), alpha (8–13 Hz), beta (13–30 Hz), and gamma (30–80 Hz; Figure 1B). For gamma, 47.5–52.5 Hz was excluded to mitigate UK powerline artifacts. Absolute band-power estimates were scaled by the total power to obtain the relative contributions within each band. Normative maps were constructed from the control average relative band power within each region and frequency band and plotted[32] (Figure 1C).

## 2.5 | Abnormality mapping of patient MEG band power

For patient abnormalities in each region and frequency band, the absolute $z$-score was computed using healthy controls as a baseline (Figure 2A). Abnormalities were computed for each patient using Equation 1, where $i$ corresponds to the region of interest, $j$ the frequency band, and $\mu_{i,j}, \sigma_{i,j}$ the mean and standard deviation of the healthy controls, respectively. To reduce the dimensionality, maximum absolute $z$-scores across frequency bands were retained for each region.

$$|z_{i,j}| = \left| \frac{x_{i,j} - \mu_{i,j}}{\sigma_{i,j}} \right| \quad (1)$$

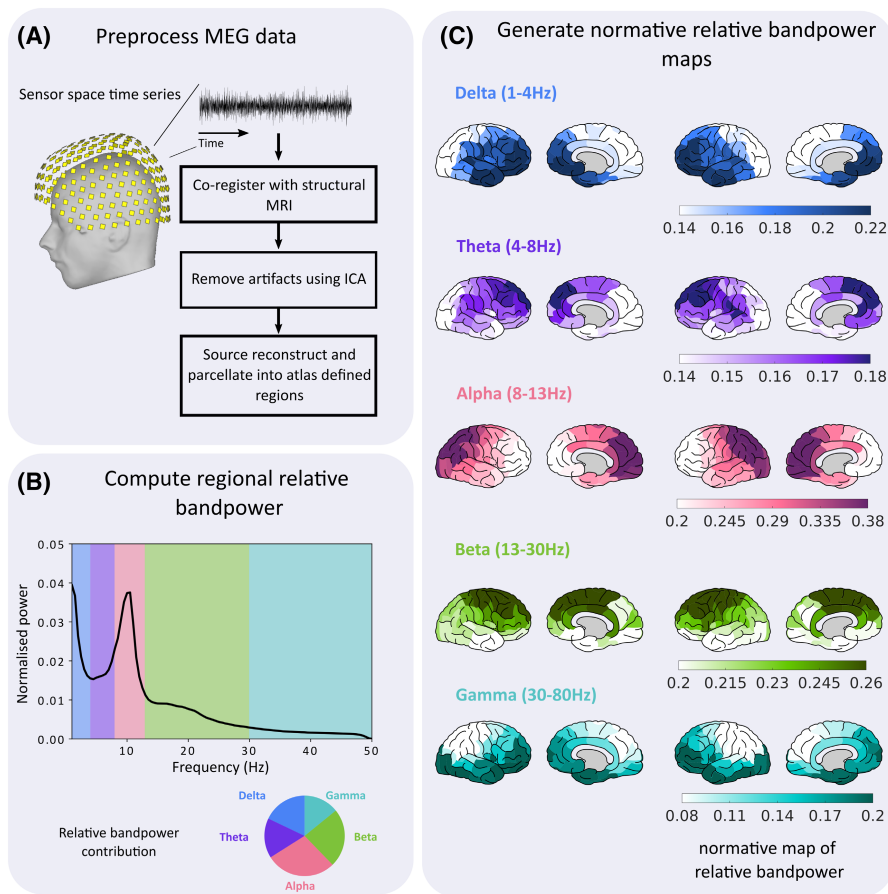

**FIGURE 1** Pre-processing pipeline for normative MEG recordings. First, (A) MEG recordings were pre-processed to remove any artifacts due to powerline interference, or cardiac and ocular events. After artifact removal, 70 s epochs of channel recordings were source reconstructed and downsampled into cortical regions based on the Lausanne parcellation. Second, (B) regional power spectral densities were computed for each patient. The relative band-power contribution was calculated for five commonly used frequency bands: delta (1–4 Hz), theta (4–8 Hz), alpha (8–13 Hz), beta (13–30 Hz), and gamma (30–80 Hz). For gamma 47.5–52.5 Hz were discarded to minimize any residual effects due to powerline artifacts. Note that the *x*-axis terminates at 50 Hz for illustrative purposes only. Finally, (C) regional band-power contributions were averaged across the cohort to create normative maps of relative band power for each of the five frequency bands.

## 2.6 | Mechanisms of surgery failure

We proposed three markers for mechanisms of surgical failure in patients with neocortical epilepsy. Each marker was quantified using the patient-specific band-power abnormality maps. Our first mechanism is failing to target abnormalities (Figure 2B). That is, if the resection area is normal (i.e., similar to the control data) we hypothesized a poor surgical outcome. We quantify this mechanism using the mean abnormality of the resection ($MA_R$)—the mean absolute *z*-score of the resected regions.

Our second mechanism of postoperative seizure recurrence is partial resection of the epileptogenic zone. Abnormal tissue may be resected, but if other more abnormal tissue is spared, an unfavorable surgical outcome may be more likely (Figure 2C). To quantify this we use the distinguishability measure $D_{RS}$,[33] which is synonymous to the area under the receiver-operating characteristic (ROC) curve (AUC). The $D_{RS}$ measure compares the rank order of abnormality of resected regions to the abnormality of the spared regions. Ranging between 0 and 1, $D_{RS}$ scores closer to 0 indicate that the most abnormal regions are resected. Conversely, values close to 1 indicate that the most abnormal regions are spared. $D_{RS}$ scores close to 0.5 correspond to chance.

Finally, we propose that insufficiently impacting the overall cortical abnormality during resection may contribute to surgical failure (Figure 2D). If high-magnitude abnormalities remain across the cortex, it is unlikely that a localized resection of abnormal tissue, even the most abnormal tissue, will sufficiently alter the overall global abnormality. We use the abnormality contribution of resected tissue on the global abnormality ($AC_R$) as a marker for this mechanism. This marker corresponds to the proportion resected tissue contributes to the overall absolute *z*-score, with overall absolute *z*-score defined as the sum of abnormalities across all cortical regions The $AC_R$ is bound between 0% and 100%, with larger values indicating that the resection more heavily contributes to the global abnormality.

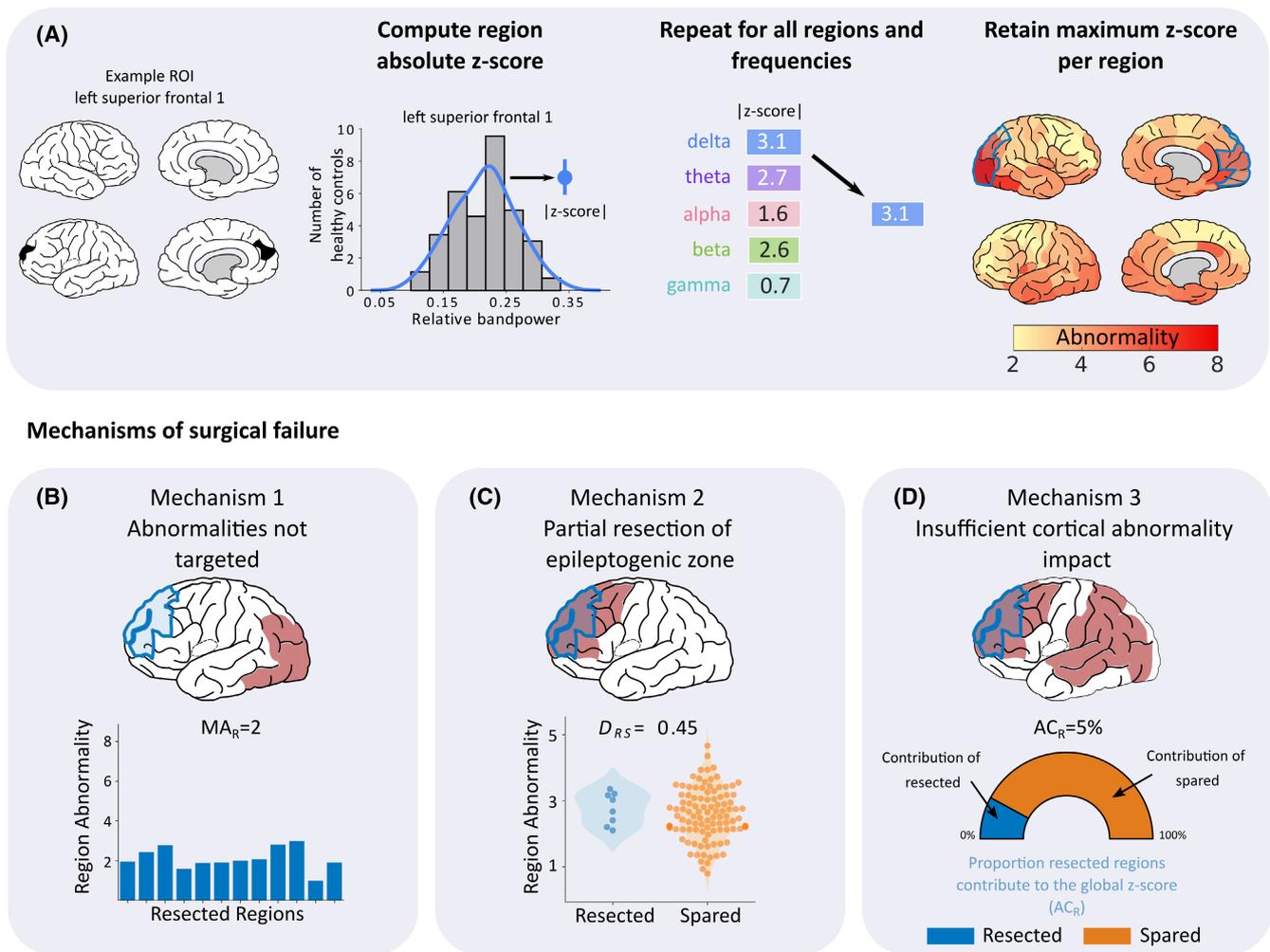

**FIGURE 2** Processing pipeline for patient abnormality mapping and derivation of the three surgical failure mechanisms. (A) Patient abnormality maps were generated using control data as a baseline (top panel). For each region and frequency, the band-power contribution was $z$-scored using control data as a reference. Within each region, frequency abnormality dimensions were reduced by retaining only the maximum absolute abnormality. This process was repeated for all regions to create patient-specific maps of band-power abnormalities. Three mechanisms were proposed that could underlie surgical failure (bottom row). (B) First, failure to resect abnormalities could lead to poor surgical outcome. This mechanism was quantified using the average abnormality of the resected tissue ($MA_R$), with lower values hypothesized to cause surgical failure. (C) Second, we hypothesized that a partial resection to the epileptogenic zone regardless of whether abnormalities are resected, could lead to surgical failure. To quantify this mechanism, we used distinguishability between resected and spared tissue ($D_{RS}$) scores to measure the separability of the resected and spared tissue based on abnormality values. $D_{RS}$ is identical to the area under the curve (AUC). Scores close to 0 indicate that the abnormalities of resected tissue are greater in magnitude than abnormalities of spared tissue. Conversely, $D_{RS}$ scores close to 1 indicate that the spared tissue is more abnormal relative to the resected tissue. (D) Finally, insufficient cortical abnormality impact could also contribute to a poor surgical outcome. We quantified this mechanism using $AC_R$, defined as the abnormality contribution of resected tissue, relative to the global abnormality. If abnormalities are widespread, the overall cortical abnormality is unlikely to be sufficiently altered by a localized resection, potentially leading to surgical failure.

## 2.7 | Statistical testing

### 2.7.1 | Surgical outcome discriminability of surgical failure mechanisms

To assess if our three proposed mechanisms explain surgery outcome (ILAE 1 vs ILAE 2+) we computed AUC values as a nonparametric measure of effect size. We hypothesized that the resections of ILAE 1 patients would be more abnormal than resections in ILAE 2+ patients. Furthermore, we hypothesized that patients with favorable outcomes would have lower $D_{RS}$ scores and higher $AC_R$ scores in comparison to poor outcome patients. Corresponding $p$-values were estimated using the Mann-Whitney $U$ test. One-tailed tests were performed as clear hypotheses of direction for each mechanism are provided.

We postulated that the $MA_R$ and $D_{RS}$ of seizure-free patients differed significantly from preconceived thresholds using one-tailed, one-sample Wilcoxon signed-rank tests. For $D_{RS}$, we hypothesized that seizure-free patients were significantly less than chance ($D_{RS} < 0.5$). In addition, we hypothesized that the $MA_R$ was significantly greater than health in seizure-free patients. To test this hypothesis, a threshold was set at |z-score| = 2.6 based on the results of a simulation. Five samples were drawn randomly from a normal distribution to simulate relative band power abnormality for each frequency band within a cortical region. We repeated this 100 000 times, retaining the maximum absolute z-score at each iteration. From these 100 000 iterations, a threshold of 2.6 corresponds to the 5% significance level.

### 2.7.2 | Simultaneous analysis of surgical failure mechanisms

Because surgical failure is multifactorial, analysis of each mechanism individually may underperform when identifying poor surgical candidates. To unify all mechanisms into a single analysis, the optimal surgical outcome separability threshold for each was calculated. This was defined as the optimal point on the ROC curve for each of the three measures that maximizes the true positive rate, while simultaneously minimizing the false positive rate.

## 3 | RESULTS

### 3.1 | Markers of surgical failure mechanisms differ across poor-outcome patients

We proposed three intuitive mechanisms that could contribute to surgical failure. Each mechanism was quantified using patient-specific band-power abnormality maps. To illustrate the importance of each mechanism, and their association with surgical failure, we provide examples using three patients with unsuccessful surgical interventions (Figure 3).

We first postulated that failing to resect abnormal tissue is related to a poor surgical outcome. For patient 1 (Figure 3, left column) the mean abnormality of the resection is low ($MA_R$ = 1.16), suggesting that most of the resected tissue is normal. Indeed, the possibility of mislocalization is apparent upon visual inspection of the corresponding abnormality map (Figure 3, top row, left panel). Strong abnormalities are present in the left frontal lobe only. For this reason, we hypothesized that the patient's right parietal lobe resection would *not* lead to postoperative seizure freedom. Conversely, patients 2 and 3 had favorable $MA_R$ scores, with values of 2.96 and 5.36, respectively, suggesting that their resection *did* target abnormal—potentially epileptogenic—tissue.

Second, we hypothesized that regardless of whether the epileptogenic zone was localized, surgical outcome would be poor if the epileptogenic zone was only partially resected and other more abnormal regions remained. We propose that sparing the *most* abnormal tissue, regardless of resection abnormality, as a marker for the second mechanism of surgical failure. The marker used for the second mechanism of surgical failure is the metric $D_{RS}$, which quantifies if the spared tissue is more abnormal than the resected tissue. Values close to 1 suggest that abnormalities were spared by surgery. For patient 2 $D_{RS}$ measured 0.73, which suggests that although possible epileptogenic abnormalities were targeted, some remaining epileptogenic tissue may have been spared. Thus we hypothesize a poor surgical outcome for the second patient.

Patient 3 demonstrated favorable markers for the first two proposed mechanisms (Figure 3, right column), with an $MA_R$ of 5.36 and $D_{RS}$ of 0.13. Based on these findings, one may expect a good outcome. However, the markers for the targeting of abnormalities and partial resection of the epileptogenic zone both fail to directly account for abnormalities beyond the resection. If abnormalities are widespread across the cortex, it is unlikely that resecting the most abnormal tissue will completely suppress seizures. Our third marker of surgical failure, $AC_R$, accounts for the abnormalities in spared tissue by quantifying the proportion that resected abnormalities contributed to the overall global abnormality. For patient 3, the $AC_R$ highlights that only 6% of the global abnormality was resected. The low contribution of the resection to the overall abnormality leads us to suggest that the localized resection was insufficient, thus leading to a poor surgical outcome.

In contrast to the non–seizure-free patients in Figures 3 and 4 highlights the markers for a patient with good surgical outcome. Patient 4 has no markers of surgical failure. First, the resection targeted abnormal tissue with an $MA_R$ of 5.95 (Figure 4B), indicating the epileptogenic zone was likely correctly localized. Moreover, a $D_{RS}$ of 0.1 indicates that the most abnormal tissue was removed, suggesting that the whole epileptogenic zone may have been resected (Figure 4C). Finally, we observed that the localized resection had a strong impact on the global abnormality, with an $AC_R$ of 19% (Figure 4D). Altogether, the results for this patient would indeed suggest a good surgical outcome based on the chosen resection site.

### 3.2 | Markers of surgical failure mechanisms discriminate outcome groups

We applied markers of the three surgical failure mechanisms to the entire cohort of 32 patients with neocortical epilepsy. Each marker was derived using patient abnormality maps with healthy controls as a baseline. Cohort-wide associations between our markers of surgical failure and surgery outcome are reported in Figure 5.

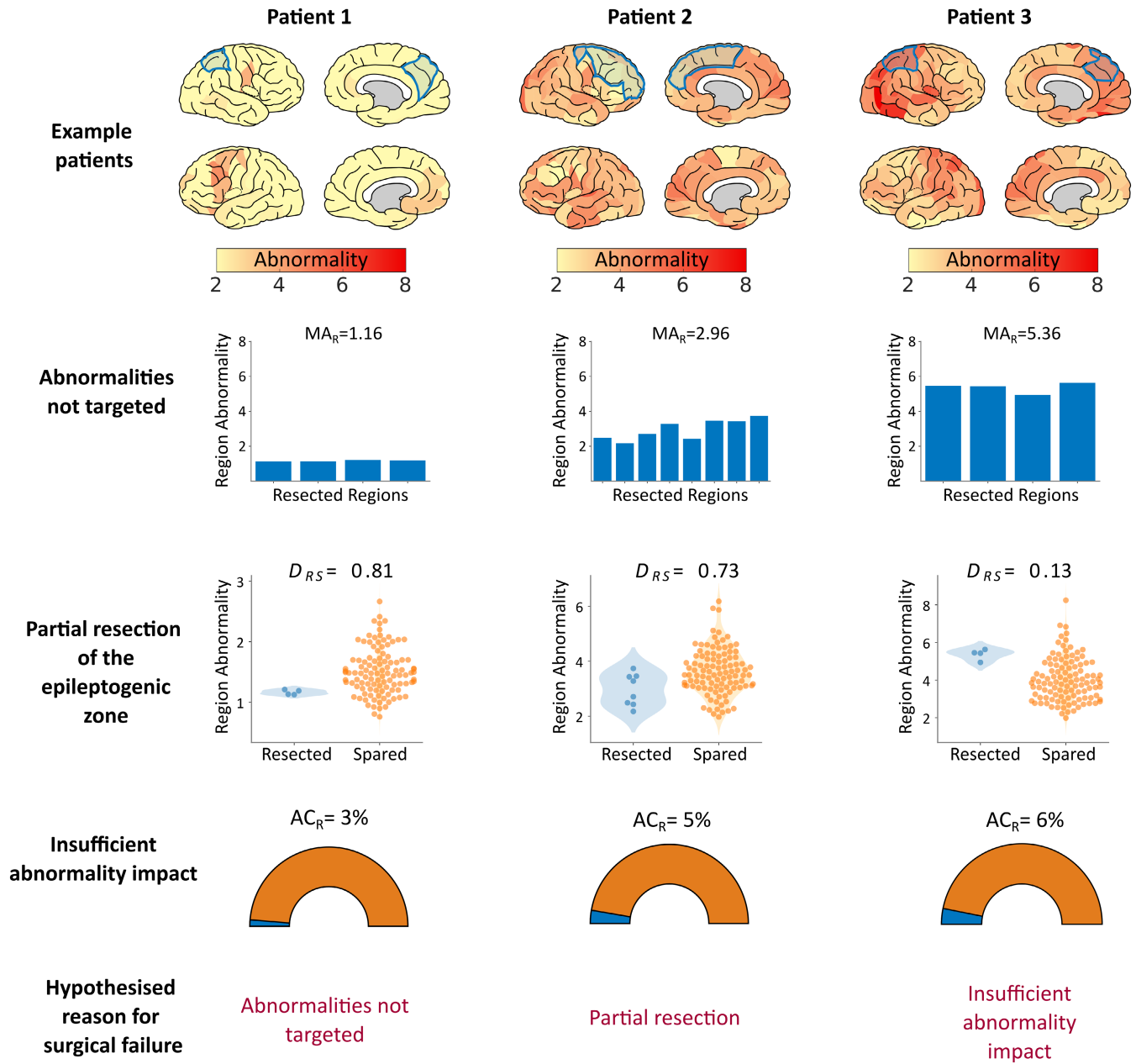

**FIGURE 3** Mechanism results for three example poor outcome (ILAE 2+) patients The top row shows patient cortical abnormality maps, with subsequent resections in blue. Lower rows show the markers for each of the three proposed mechanisms of surgical failure. Each column corresponds to a different patient, illustrating the variety of pathways (consisting of one or more mechanisms) that could lead to a surgical failure. For patient 1 (left column), it is evident that abnormalities were not resected, with an average resection abnormality ($MA_R$) of 1.16. As such, it is likely that the failure to resect abnormalities contributed to the patient's poor surgical outcome. For patient 2 (center column), abnormalities were resected but the most abnormal tissue was spared (distinguishability between resected and spared, $D_{RS} = 0.73$), suggesting that the epileptogenic zone may have been partially resected. The corresponding abnormality map illustrates that widespread abnormalities were present, a confounder that may have led to the failed surgical intervention. Finally, for patient 3 (right column), although some of the most abnormal cortical tissue was resected, other abnormalities were widespread. As such, the global abnormality was likely not sufficiently altered (abnormality contribution of the resected tissue, $AC_R = 6\%$). All three patients fail at least one of our proposed mechanisms.

For the $MA_R$ scores, which correspond to the targeting of abnormal tissue (Figure 5A), we see that the $MA_R$ scores of seizure-free (ILAE 1) patients are larger than those of non–seizure-free (ILAE 2+) patients (AUC = 0.80, $p = .003$). This result highlights that resections with stronger abnormalities are associated with surgical success. In addition, the $MA_R$ scores for seizure-free patients differed from health ($MA_R > 2.6$, $W = 70.0$, $p = .006$). Meanwhile, our marker for the partial resection of the epileptogenic zone, $D_{RS}$, was larger in patients with poor surgical outcome (AUC = 0.68, $p = .053$). That is, in poor outcome patients the most abnormal cortical tissue was typically spared rather than resected (Figure 5B). Moreover, $D_{RS}$ scores for good outcome patients were less than chance ($W = 10.0$, $p = .010$), meaning that resected regions were more abnormal than spared. Finally, our third marker corresponding to insufficiently altering the global abnormality ($AC_R$) was larger in good outcome patients. This marker quantifies the impact that the resection has on the overall abnormality load (Figure 5C). This effect is in the hypothesized direction, being a larger impact in seizure-free patients (AUC = 0.64, $p = .096$). These results show that, as in our example patients, our markers of surgical failure mechanisms may identify patients with poor surgical outcomes across the cohort studied.

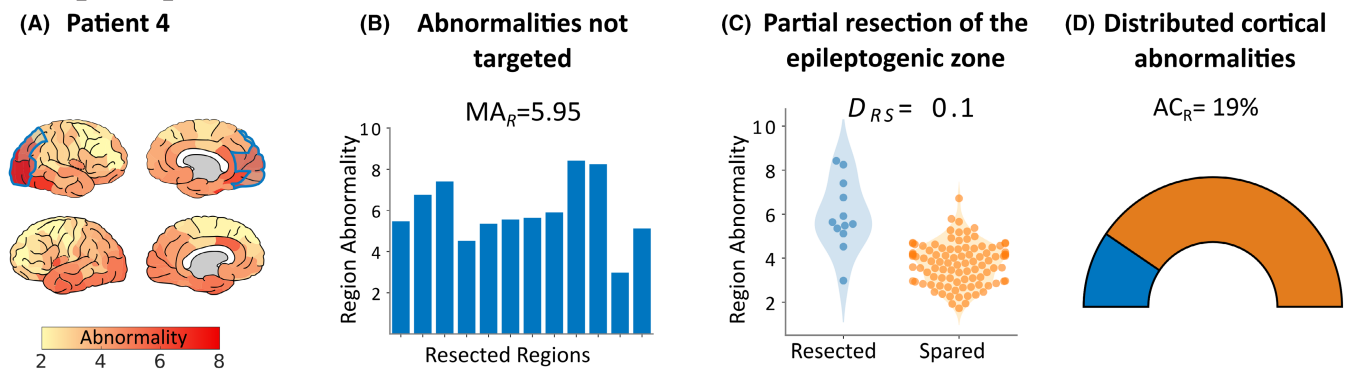

**FIGURE 4** Mechanism results for an example good outcome patient. (A) Patient 4 underwent resective surgery, achieving total seizure suppression 1 year postoperatively. For all three mechanisms, patient 4 passes criteria to suggest surgical success. (B) First, abnormal tissue has been targeted within the resection cavity. (C) In addition, it is clear that the most abnormal cortical tissue has been targeted. (D) Finally the abnormality contribution of the resected tissue ($AC_R$) suggests that the overall cortical abnormality may have been sufficiently altered after resection. Together, these results based on pre-operative interictal MEG recordings would lead us to hypothesize that the proposed resection zone for this patient is sufficient to completely suppress seizures.

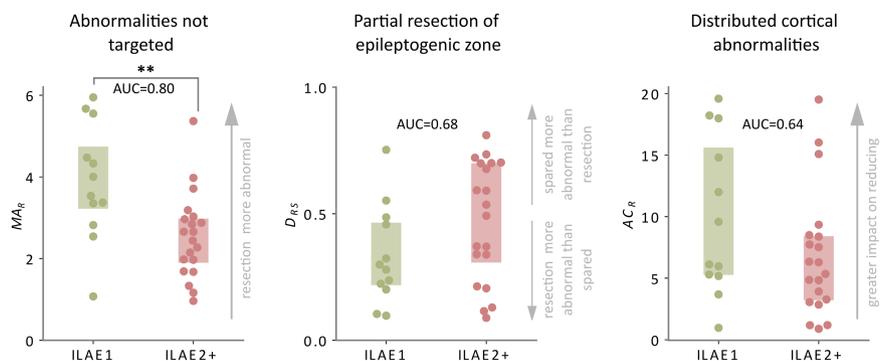

**FIGURE 5** Surgical outcome separability of each mechanism at the group level. Boxplots illustrate how well each mechanism discriminates surgical outcome groups. Each datapoint represents an individual patient. Boxes extend from the 25%–75% of the data. For each mechanism, the area under the curve (AUC) is calculated along with corresponding $p$-value using a one-tailed Mann-Whitney $U$ test. Good outcome patients (ILAE 1) are depicted in green, and bad outcome (ILAE 2+) in red. (**) corresponds to statistical significance at the 1% level.

## 3.3 | Simultaneous analysis of surgical failure mechanisms relates to postsurgical outcome

Epilepsy is heterogeneous, with a wide range of mechanisms by which seizures may originate and manifest. It is, therefore, reasonable to assume that there are multiple mechanisms that render treatment ineffective, with different mechanisms in different patients. Analyzing each mechanism independently may, therefore, underperform compared to a unified analysis.

The three proposed mechanisms of surgical outcome failure provided complementary information (Figure S2).

Given this complementary information, optimal thresholds to discriminate surgical outcome groups were next chosen based on a data-driven approach (see Methods section 2.7.3). A patient would be considered a poor surgical candidate if the mean abnormality of the resection ($MA_R$) was less than 3.35. For patients with partial resections of the epileptogenic zone or insufficient impact to the cortical abnormality we would expect to see a $D_{RS} > 0.34$ or $AC_R < 9.58\%$, respectively. Some 17 of 20 (85%) non–seizure-free patients had $MA_R$ scores less than 3.35. Similarly, 85% of non–seizure-free patients had $AC_R$ values less than 9.58%. For non–seizure-free patients, 15 of 20 (75%) had $D_{RS}$ scores greater than 0.34, a marker of surgical failure due to the partial resection of the epileptogenic zone. Combining all mechanisms into a single analysis showed that 19 of the 20 (95%) non–seizure-free patients were identified with at least one marker of surgical failure (Figure 6).

Overall, 27 patients had at least one marker of surgical failure, of which 70.37% went on to have poor surgical outcomes. Moreover, 90% and 60% of poor outcome patients had at least two and three markers of surgical failure, respectively. Conversely, seizure-free patients had at least one, two, and three markers of surgical failure with rates of 66%, 25%, and 16%, respectively. Using the number of markers of surgical failure, the unified analysis distinguishes surgical outcome groups well (AUC = 0.82, $p = .0008$). For patients for which our confidence of postoperative seizure freedom was highest (i.e., no markers of surgical failure), we were correct in four of five, with the remaining patient later recovering from year 2. To summarize, if a patient failed at least one marker, there was a 70.37% chance of poor outcome. In contrast, if a patient passed all three markers, there was an 80% chance of good outcome.

## 4 | DISCUSSION

Different patients may have different reasons for postsurgical seizure recurrence. Here we proposed three mechanisms relating to epilepsy surgical failure. Markers for each mechanism were quantified using patient-specific band-power abnormality maps from interictal MEG recordings, acquired during pre-surgical evaluation. First, we demonstrated that each marker of surgical failure relates to surgical outcome in the hypothesized direction in four example case studies. Second, we showed that these findings held across our cohort as a whole, with the resection of abnormalities ($MA_R$) significantly separating outcome groups. Finally, by identifying optimal thresholds of separability, we showed that 95% of non–seizure-free patients had at least one marker of surgical failure, suggesting our mechanisms could aid pre-surgical evaluation.

Both intuitive, and easy to quantify, each mechanism provides important insights into why some patients fail to have complete suppression of seizures postoperatively. We hypothesized that not targeting abnormal, possibly epileptogenic, tissue would result in surgical failure. This mechanism builds on extensive literature relating structural abnormalities of the resected tissue to surgical outcome.[13–16,22] We quantified this mechanism as a single value per patient using the mean resection abnormality (or $MA_R$). Our results suggest that the resection of abnormalities ($MA_R$) is indeed crucial for postoperative seizure control, and that normative maps can be used to identify abnormalities.

It is conceivable that the resection may indeed remove abnormal tissue, but that other, even more abnormal areas remain (e.g., in eloquent cortex). Our first marker measures only properties of resected regions, disregarding the remaining tissue. We, therefore, postulated that patients with only partial resections of abnormalities would achieve only partial seizure control. We quantified the partial resection using a second marker, $D_{RS}$,[33,34] a measure that captures whether resected tissue is more abnormal than spared. Consistent with previous work, we show that this marker relates to outcome in the hypothesized direction (AUC > 0.5). Moreover, the resection of the most abnormal tissue in seizure-free patients was significantly more common than chance ($D_{RS}$ < 0.5), suggesting that these regions were not targeted randomly during surgery, but correctly identified in seizure-free patients. In contrast, $D_{RS}$ scores for non–seizure-free patients were not significantly different than chance, indicating that the most abnormal tissue was not removed in those patients. Supporting our findings, Englot et al.[3] reported that surgical intervention did not completely suppress seizures in 72% of the non–seizure-free patients because the epileptogenic zone was only partially resected. Similarly, Kim et al.[35] also reported that complete resection of the ictal-onset areas, identified by frequent interictal spiking during an intracranial electroencephalography (iEEG) study, related to better surgical outcome.

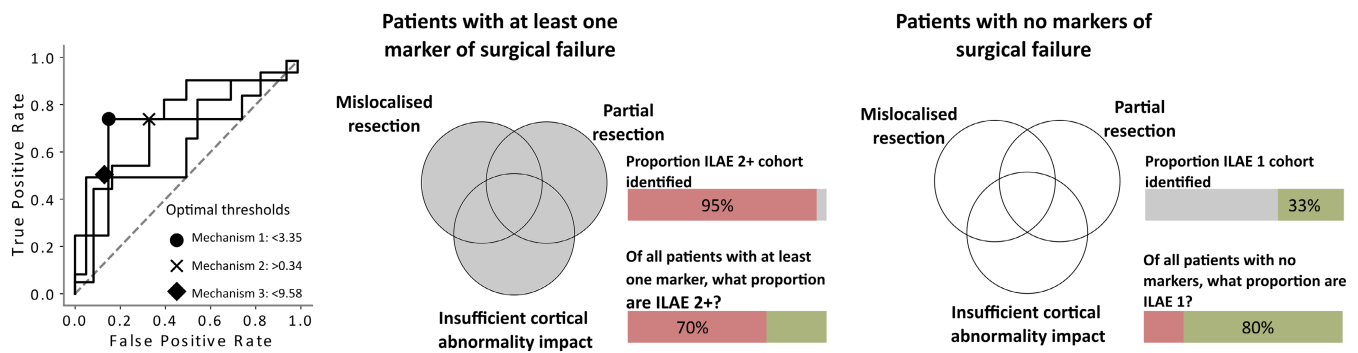

**FIGURE 6** Harmonization of all three mechanisms into a single analysis. Receiver-operating characteristic (ROC) curves (left) for each mechanism are used to identify the optimal threshold that discriminates the two surgical outcome groups. Optimal thresholds are selected at the point that maximizes the true positive rate while simultaneously minimizing the false-positive rate. Patients are filtered into sub groups based on those that fail to meet the criteria for at least one mechanism (center) and those that pass all three mechanisms (right). Shaded areas (gray) of Venn diagrams illustrate the criteria used to filter the cohort. Specificity corresponds to the proportion of good or bad outcome patients identified relative to all patients identified for the corresponding criteria.

Our third mechanism hypothesized that if strong cortical abnormalities existed beyond the proposed resection site, a localized resection may not fully suppress seizures postsurgically. Widespread abnormalities have been reported previously using diffusion MRI data in epilepsy,[36,37] with some studies attributing widespread abnormalities to an underlying epileptogenic network facilitating the spread of seizures to distant regions.[7,38] Our marker of surgical failure may reflect this epileptogenic network. We note that although our marker may be sensitive to epileptogenic abnormalities it may not be specific and may detect other non-epileptogenic abnormalities. Careful interpretation using other clinical information would be required to assess which abnormalities are associated with epileptogenic tissue. Together our three mechanisms account for abnormalities within both the resected and spared tissue.

In isolation, each mechanism discriminates surgical outcome groups in the hypothesized directions. However, surgery failures may occur for different reasons,[2] and so it is unlikely a single mechanism will fully distinguish surgical outcome groups. Our formalism of these mechanisms allowed us to demonstrate that 95% of poor outcome patients presented with at least one marker of surgical failure, 90% with at least two markers, and 60% with markers of surgical failure for all three mechanisms. Overall, a combination of all three mechanisms discriminated surgical outcome groups well (AUC = 0.82, $p$ = .0008). Fourteen patients were identified with three markers of surgical failure and thus would be hypothesized to be poor surgical candidates. In reality, 86% of those patients with three markers of surgical failure were indeed not seizure-free postsurgically. In total, five patients had no markers of surgical failure based on our proposed mechanisms, thus suggesting good outcomes for all five. Indeed, four of those were seizure-free, with the remaining patient ILAE 3 at 12 months. However, that patient experienced only one seizure while in the process of reducing their medication during their first months postsurgery. Although having an initially poor outcome due to the solitary seizure in year 1, the patient was seizure-free for all subsequent years (over 5 years to date). Our results suggest that the analysis of these surgical failure mechanisms, using a multimetric framework, provides robust predictions of surgical outcome of use during pre-surgical evaluation.

Although we proposed three mechanisms of surgical failure, others may exist, including the presence of multiple epileptogenic foci, or the development of new epileptogenic zones postoperatively. To assess whether multiple foci are present, spatial statistics could be used[39,40] to investigate spatial organization of abnormalities and their proximity to the resection. Such approaches could be useful to identify multifocal abnormalities, which may be a contraindication for surgery. To investigate the development of new epileptogenic zones postoperatively, network-based approaches could be beneficial. Recently, it was proposed that there exists widespread postsurgical functional abnormalities that were not present pre-surgically.[4] There, the authors hypothesized that postsurgical evolution of the epileptogenic process may have contributed to continued seizures in poor outcome patients. In addition, a longitudinal study of 48 patients following temporal lobe resections revealed greater white matter postoperative changes related to better outcomes.[41] Future studies could develop markers pertaining to new mechanisms of surgical failure and incorporate them with our mechanisms.

During pre-surgical evaluation, interictal recordings are used to identify and localize interictal spikes as markers of the epileptogenic zone.[42] However, the origin and pathological nature of spikes has been questioned, with some studies suggesting a protective effect,[43] and others suggesting that their resection does not improve outcomes.[20] Given the high expense of MEG, the clinical use of interictal MEG recordings is less common during pre-surgical evaluation compared to some other modalities, and typically reserved for patients with difficult to localize seizure foci. We leverage the usually discarded interictal data, demonstrating that even seemingly normal recordings can contain clinically valuable localization information. Beyond application to traditional MEG data, future studies could apply our markers of surgical failure to scalp EEG, or less-expensive portable optically pumped magnetometer-based MEG.

Limitations of this work exist. First, although a large control cohort was used, the patient sample size used in this study is limited, precluding the use of some machine learning techniques. Instead, we proposed intuitive and hypothesis-driven mechanisms, which should generalize across patient cohorts. A second limitation is the poor signal-to-noise ratio of resting state MEG for subcortical structures such as the hippocampus and amygdala, with challenges due to the inverse problem. This is particularly problematic for the resting state low-amplitude activity studied here. Due to these factors, we excluded deep brain structures from our analysis, and include patients with hippocampal resections only as supplementary material. Finally, our normative data were acquired at a site different from that of our patients. However, both sites used the same scanners, and we expect that our use of relative band power largely mitigates the site differences. Furthermore, although our normative data are used as baseline, all statistical comparisons regarding surgical outcome were between patients from the same site. It should be noted that normative maps can vary with age.[44,45] Future studies wishing to analyze pediatric patients should ideally recompute the normative maps using age-matched healthy pediatric controls. In this study we retrospectively used a mask of tissue that was resected. A future prospective application could use a mask of the intended resection. Furthermore, our abnormality maps could be used prospectively to guide electrode implantation for those patients undergoing subsequent iEEG recordings, and the location and extent of the resections.

Surgical resection of the epileptogenic zone is a treatment option for patients with refractory focal epilepsy. At present, it is not fully understood why surgical intervention in some patients is unsuccessful, although there may be different reasons for different patients. Herein, we proposed three mechanisms that relate to surgical failure. These mechanisms could aid clinicians during the pre-surgical evaluation of patients, particularly in patients with difficult-to-treat neocortical epilepsy.

## AUTHOR CONTRIBUTIONS

Thomas W. Owen, Yujiang Wang, and Peter N. Taylor contributed to the conception and design of the study. Thomas W. Owen, Gabrielle M. Schroeder, Vytene Janiukstyte, Gerard R. Hall, Andrew McEvoy, Anna Miserocchi, Jane de Tisi, John S. Duncan, Fergus Rugg-Gunn, Yujiang Wang, and Peter N. Taylor, contributed to the acquisition and analysis of data. Thomas W. Owen, Gabrielle M. Schroeder, and Peter N. Taylor contributed to drafting of the text and preparing the figures.

## ACKNOWLEDGMENTS

We thank members of the Computational Neurology, Neuroscience & Psychiatry Lab (www.cnnp-lab.com) for discussions on the analysis and manuscript. The normative data collection was supported by an MRC UK MEG Partnership Grant, MR/K005464/1.

## FUNDING INFORMATION

T.O. was supported by the Centre for Doctoral Training in Cloud Computing for Big Data (EP/L015358/1). P.N.T. and Y.W. are both supported by UKRI Future Leaders Fellowships (MR/T04294X/1, MR/V026569/1). The normative data collection was supported by an MRC UK MEG Partnership Grant, MR/K005464/1. J.S.D is supported by the Wellcome Trust Innovation grant 218380. J.S.D and J.dT are supported by the NIHR UCLH/UCL Biomedical Research Centre.

## CONFLICT OF INTEREST

No relevant conflicts of interest are reported.

## ORCID

*Gabrielle M. Schroeder*  https://orcid.org/0000-0003-2278-5227
*Gerard R. Hall*  https://orcid.org/0000-0002-5212-7850
*Peter N. Taylor*  https://orcid.org/0000-0003-2144-9838

## SUPPORTING INFORMATION

Additional supporting information can be found online in the Supporting Information section at the end of this article.